\documentclass[12pt,a4paper]{article}
\usepackage{graphicx,epsfig}
\usepackage{color}
\usepackage{amsmath}
\usepackage{amsfonts}
\usepackage{amssymb}
\usepackage{cite}
\usepackage{ifpdf}
\newcommand{\be}{\begin{equation}}
\newcommand{\ee}{\end{equation}}
\newcommand{\bea}{\begin{eqnarray}}
\newcommand{\eea}{\end{eqnarray}}
\newcommand{\ba}{\begin{array}}
\newcommand{\ea}{\end{array}}

\newcommand{\cB}{{\cal B}}
\newcommand{\cO}{{\cal O}}
\newcommand{\cF}{{\cal F}}

\newcommand{\no}{\nonumber}

\newcommand{\lsim}{\stackrel{<}{_\sim}}
\newcommand{\gsim}{\stackrel{>}{_\sim}}

\begin{document}

\begin{center}
{\Large \bf Drell-Yan production of Heavy Vectors \\[0.2 cm] in Higgsless models}\vskip 1.0cm
{\large  Oscar Cat\`a, Gino Isidori, Jernej~F.~Kamenik} \\[0.5 cm]
{\it INFN, Laboratori Nazionali di Frascati, Via E. Fermi 40
I-00044 Frascati, Italy}\\[1.0 cm]
{\bf Abstract}
\begin{quote}
We study the Drell-Yan production of heavy vector and axial-vector 
states of generic Higgsless models at hadron 
colliders. We analyse in particular 
the $\ell^+\ell^-$, $WZ$, and three SM gauge boson 
final states. In the  $\ell^+\ell^-$ case we show how 
present Tevatron data restricts the allowed parameter 
space of these models. The two and three gauge boson
final states (especially $WZ$, $WWZ$, and $WZZ$) 
are particularly interesting in view of the LHC,
especially for light axial-vector masses, and could shed 
more light on the role of spin-1 resonances in the 
electroweak precision tests. 
\end{quote}
\end{center}
\vskip 1.0cm

\section{Introduction} 

While the evidences in favour of the spontaneous 
breaking of the electroweak group 
are very strong, the fact that this breaking occurs via 
a single fundamental Higgs field, with a non-trivial vacuum
expectation value, is far from being clearly established. 
A fundamental Higgs boson is certainly the most economical
way to explain this spontaneous breaking, and a light Higgs mass 
($m_{h}\approx $~100~GeV) is also an efficient way to account 
for all the existing electroweak precision tests. However, 
the strong sensitivity of $m_{h}$ to short-distance scales 
poses a serious naturalness problem to this view and 
motivates the search for alternative symmetry-breaking mechanisms. 
An interesting alternative is that of {\em Higgsless} models,
or the wide class of theories (see 
e.g.~Ref.~\cite{Casalbuoni:1985kq,Csaki:2003dt,Nomura:2003du,
Barbieri:2003pr,Foadi:2003xa,Georgi:2004iy})
where the  $SU(2)_L \times U(1)_Y \to U(1)_Q$ 
breaking is generated by some new strong dynamics 
above the Fermi scale ($v=(\sqrt{2}G_{F})^{-1/2}\simeq246$~GeV).

A general feature of Higgsless models is the appearance of 
new spin-1 states that replace the Higgs boson in keeping 
perturbative unitarity up to a few TeV~\cite{Chivukula:2003kq,Bagger:1993zf}.
These states are 
the lightest non-standard particles and should provide 
the first clue of such models at high-energy colliders. 
The phenomenology of heavy vectors at Tevatron and the LHC 
has been discussed by various authors (for recent 
analyses see e.g.~\cite{He:2007ge,Accomando:2008jh,Belyaev:2008yj}).
However, most of the existing analyses are based on specific
dynamical assumptions, e.g., considering these vector 
states as the massive gauge bosons of a hidden 
local symmetry. As recently discussed in~\cite{Barbieri:2008cc}, 
these assumptions may be too restrictive for generic models with 
strong dynamics at the TeV scale, and only 
going beyond these assumptions the sole exchange of 
heavy vectors can provide a successful description of 
electroweak precision observables (EWPO). 

The purpose of this paper is to study the Drell-Yan 
production (and subsequent decays) of spin-1 states at hadron 
colliders, based on the following rather general dynamical assumptions:
\begin{itemize}
\item The new strong dynamics is invariant under 
a global chiral symmetry $G=SU(2)_L \times SU(2)_R$, broken 
spontaneously into $H=SU(2)_{L+R}$ (the {\em custodial symmetry} of the SM Higgs potential),
and under a discrete parity symmetry ($P:\ SU(2)_L \leftrightarrow SU(2)_R$). 
\item A pair of vector ($V$) and axial-vector ($A$) states,
belonging to the adjoint representation of $H$, are the only new {\em light} 
dynamical degrees of freedom below a cut-off scale 
$\Lambda \sim (2\div3)$~TeV. 
\item The exchange of the $V$ field ensures the 
tree-level unitarity of $W W \to W W$ scattering 
up to the cut-off.
\end{itemize}
Employing these general assumptions, we consider an effective theory 
valid below the cut-off scale $\Lambda$ written in terms of the 
Goldstone bosons of the $G/H$ breaking (which give rise to 
the longitudinal components of the $W$ and $Z$ fields) and the 
new spin-1 states. As in~\cite{Barbieri:2008cc},
we describe the latter using the antisymmetric 
tensor formalism of Ref.~\cite{Ecker:1989yg}, 
and restrict the parameter space of the model 
imposing the EWPO constraints.

Within this framework we study the Drell-Yan production
of the $V$ and $A$ states, and their subsequent decays into 
$\ell^+\ell^-$ pairs, two and three SM gauge bosons. 
Even after imposing the EWPO constraints a wide range of masses and couplings for the  $V$ 
and $A$ states is allowed. We show that a significant reduction 
of the allowed parameter space is obtained using the 
recent $p \bar p \to e^+e^-$ data from Tevatron. As far as the 
future prospects of discovering such states at the LHC are concerned,
we do not present a general scan over the allowed 
parameter space. Instead we focus our attention on: (i) identifying 
interesting experimental signatures at the early stage of the LHC; and (ii)
identifying the main differences with respect to more 
conventional Higgsless models. 

The paper is organised as follows: in Section~\ref{sect:compo} 
we present the Lagrangian of the model. Analytic expressions 
for the decay widths and the Drell-Yan cross sections (in some 
simplifying limits) of $V$ and $A$ states are presented in 
Section~\ref{sect:widths} and~\ref{sect:cross},
respectively. The numerical analysis is presented in 
Section~\ref{sect:num} and the results 
are summarised in the Conclusions.

\section{The Lagrangian}
\label{sect:compo}

The starting point is the usual lowest-order chiral Lagrangian for the
$G/H$ Goldstone boson fields
\begin{equation}
\mathcal{L}_{\chi}^{(2)}(U)=\frac{v^{2}}{4}\langle D_{\mu}U(D^{\mu}%
U)^{\dagger}\rangle\, , \label{eq:L2}%
\end{equation}                
where $v=(\sqrt{2}G_{F})^{-1/2}\simeq246$~GeV,
\begin{align}
&  U=u^2=e^{i2\hat{\pi}/v},\qquad\hat{\pi}=T^{a}\pi^{a}=\frac{1}{\sqrt{2}}\left[
\begin{array}
[c]{cc}%
\frac{\pi^{0}}{\sqrt{2}} & \pi^{+}\nonumber\\
\pi^{-} & -\frac{\pi^{0}}{\sqrt{2}}%
\end{array}
\right]  ~,\qquad T^{a}=\frac{1}{2}\sigma^{a},\nonumber\\
&  D_{\mu}U=\partial_{\mu}U-i\hat{B}_{\mu}U+iU\hat{W}_{\mu}~,\qquad\hat
{W}_{\mu}=gT^{a}W_{\mu}^{a}~,\qquad\hat{B}_{\mu}=g^{\prime}T^{3}B_{\mu}~,
\label{def}%
\end{align}
and $\langle\rangle$ denotes the trace of a $2\times2$ matrix. 
The transformation properties of the Goldstone boson 
fields under $SU(2)_{L}\times SU(2)_{R}$ are defined by 
\begin{equation}
u\rightarrow g_{R}uh^{\dagger}=hug_{L}^{\dagger}~,\qquad U\rightarrow g_{R}Ug_{L}^{\dagger}~,
\end{equation}
where $h=h(u,g_{L},g_{R})$ is an element of  $SU(2)_{L+R}$.

Following Ref.~\cite{Ecker:1989yg}, we describe the heavy spin-1 states by
means of antisymmetric tensors. We consider two phenomenologically 
relevant vector states with opposite parity, $A^{\mu\nu}[1^{++}]$ and $V^{\mu\nu}[1^{--}]$, both belonging to the adjoint representation of
$SU(2)_{L+R}$:
\begin{equation}
R^{\mu\nu}\rightarrow hR^{\mu\nu}h^{\dagger}~,\qquad R^{\mu\nu}=A^{\mu\nu
},\ V^{\mu\nu}~. \label{eq:Rtr}%
\end{equation}
The kinetic term in the Lagrangian has the form:
\begin{equation}
\mathcal{L}_{\mathrm{kin}}(R^{\mu\nu})=-\frac{1}{2}\langle\nabla_{\mu}%
R^{\mu\nu}\nabla^{\sigma}R_{\sigma\nu}\rangle+\frac{1}{4}M_{R}^{2}\langle
R^{\mu\nu}R_{\mu\nu}\rangle~,
\end{equation}
with the covariant derivative
\begin{equation}
\nabla_{\mu}R=\partial_{\mu}R+[\Gamma_{\mu},R],\qquad\Gamma_{\mu}=\frac{1}%
{2}\left[  u^{\dagger}(\partial_{\mu}-i\hat{B}_{\mu})u+u(\partial_{\mu}%
-i\hat{W}_{\mu})u^{\dagger}\right]  ,\quad\Gamma_{\mu}^{\dagger}=-\Gamma_{\mu
}.
\end{equation}
The $\mathcal{O}(p^{2})$ couplings of these heavy fields to
Goldstone bosons and SM gauge fields are parametrised in terms of 
3 effective operators, defined by 
\bea
\mathcal{L}_{1V}^{(2)} &=& \frac{i}{2\sqrt{2}}G_{V}\langle V^{\mu\nu}[u_{\mu
},u_{\nu}]\rangle +\frac{1}{2\sqrt{2}}F_{V}\langle V^{\mu\nu}(u\hat{W}^{\mu\nu
}u^{\dagger}+u^{\dagger}\hat{B}^{\mu\nu}u)\rangle \no \\
&+&\frac{1}{2\sqrt{2}}F_{A}\langle A^{\mu\nu}(u\hat{W}^{\mu\nu}u^{\dagger}-u^{\dagger}\hat{B}%
^{\mu\nu}u)\rangle~, \label{eq:LV}%
\eea
where 
$u_{\mu}=iu^{\dagger}D_{\mu}Uu^{\dagger}=u_{\mu}^{\dagger}$
(such that $u_{\mu}\rightarrow hu_{\mu}h^{\dagger}$).
The 3 effective couplings, $F_{V,A}$ and $G_V$, have dimensions of mass
and, by naive dimensional analysis, are expected to be of $\cO(v)$.

As shown in Ref.~\cite{Barbieri:2008cc}, an important role in 
electroweak precision tests is also played by  $\mathcal{O}(p^{2})$  operators 
with two heavy fields. In particular, we are interested in the following effective
Lagrangian
\bea
\mathcal{L}_{2V}^{(2)} = ig_{A}\langle A^{\mu\nu}[\nabla_{\rho}V^{\rho\nu},u_{\mu}]\rangle
+ig_{V}\langle V^{\mu\nu}[\nabla_{\rho}A^{\rho\nu},u_{\mu}]\rangle~,\label{eq:L2V}
\eea
where the adimensional couplings $g_{V,A}$ are expected to be $\cO(1)$.

\section{Decay widths}
\label{sect:widths}

We compute the decay widths of the heavy vectors at tree level, expanding the matrix 
elements to first non-trivial order in $\epsilon = v^2/M_R^2$. 
For the amplitudes we are interested in, the expansion 
in $\epsilon$ is accompanied by an expansion in the gauge 
coupling, and the effective expansion parameter 
turns out to be $g^2\epsilon^2 \sim m^2_W/M^2_R$. 
This implies that neglecting higher order terms
in this expansion is usually a good approximation 
even for $M_R \sim 500$~GeV. The only exceptions are cases 
where the leading result has some anomalous suppression, 
such as the possible phase-space suppression of the 
axial decay width (see below).

On the other hand, we do not neglect terms suppressed by 
$s_W^2 \approx 0.23$, which is often not a good numerical 
approximation.
We finally assume $M_A > M_V$, as expected in all realistic models, 
such that the vector state cannot decay into the axial one.

\subsection{Vector fields}

The leading decay channel of the heavy vector is the two-body decay 
into two longitudinal SM gauge bosons. Taking into account that  
$G_{V}=\cO(v)$, the corresponding decay width is
of $\cO(M_V \epsilon^{-2})$. 
In particular, for the charged and neutral states we have 
\be
\Gamma_{V^+} \approx \Gamma^V_{WZ} =
\frac{G_{V}^{2}M_{V}^{3}}{48\pi v^{4}} 
\left[1+\cO(g^2\epsilon^2)\right]~, \qquad\quad 
\Gamma_{V^0} \approx
\Gamma^{V}_{WW} = \Gamma^V_{WZ}
\left[1+\cO(g^2\epsilon^2)\right]~. 
\ee
The two-body decays 
into fermion pairs are parametrically suppressed 
by $\cO( g^4 \epsilon^4)$ with respect to the leading mode, and thus are 
strongly suppressed:
\be
\Gamma^V_{\bar d u} = 3 \Gamma^V_{\bar \ell \nu} = 
\frac{ g^{4}F_{V}^{2}}{64\pi M_{V}}~, \qquad\quad
\Gamma^V_{\bar f f} = \frac{2 N_f \Gamma^V_{\bar \ell\nu}}{c_W^4} 
\left\{ \left[ (1-2s_W^2) T_3^f +s_W^2 Q_f \right]^2 
+ (s_W^2 Q_f)^2 \right\}~,
\label{eq:Vff}
\ee
where $N_f$ denotes the color and flavor multiplicity of the 
final state.\footnote{~For completeness, we recall that
$T_3^u=T_3^\nu=-T_3^d=-T_3^\ell=1/2$, $Q_\ell=-1$ and $Q_u=-2 Q_d=2/3$. }

\subsection{Axial fields}

In the axial case, due to $M_A>M_V$, $\mathcal{L}_{2V}^{(2)}$ allows the decay channel $A\to V+W(Z)$, which turns out to be the parametrically leading mode:
\bea
&& \Gamma^A_{V^+W^-} = \Gamma^A_{V^-W^+}=
\Gamma^A_{V^0 W^+}= \Gamma^A_{V^+ Z} \doteq  \Gamma^A_{VW}\, ,
 \no \\
&&  \Gamma^A_{VW} =
\frac{ M_A^3}{48\pi v^2} (1-r^2)^3
\left[g_A^2 (1+2 r^2) + g_V^2\left( 1 +\frac{2}{r^2}\right) + 6 g_A g_V \right]~,
\eea
where $r=M_V/M_A$.

In the case of the axial vectors, parity forbids the 
decay into two longitudinal SM gauge bosons. Therefore, the leading decay mode allowed by $\mathcal{L}_{1V}^{(2)}$ is now the channel with one transverse and one longitudinal 
gauge boson, whose corresponding decay width is of $\cO(M_A g^2)$.
In particular, for the neutral and charged channels we have:
\be
\Gamma^A_{WW} = \frac{ g^2 F_A^2 M_{A}}{192\pi v^2}~, \qquad 
\Gamma^A_{WZ} = \frac{1}{2} \Gamma^A_{WW}
\left[ 1 + \frac{(1-2s_W^2)^2}{c^2_W}\right]~, \qquad 
\Gamma^A_{W\gamma} = 2 s_W^2~\Gamma^A_{WW}~.
\ee
As anticipated, the $A\to V+W(Z)$ decay channel may be affected by a sizable kinematical suppression in the limit $r\rightarrow 1$. As a result, a safe approximation to the total width of the axial vector 
is obtained by summing the $VW$ and $WW$ final states:
\be
\Gamma_{A^+} \approx 2 \Gamma^A_{VW} + \Gamma^A_{WZ} + \Gamma^A_{W\gamma}~, 
\qquad 
\Gamma_{A^0} \approx 2 \Gamma^A_{VW} + \Gamma^A_{WW}~.
\ee
The two-body fermionic decay widths of the axial vectors 
are identical to those in Eq.~(\ref{eq:Vff}) but for 
$\cO(s^2_W)$ corrections in the neutral case:
\be
\Gamma^A_{\bar d u} = 3\Gamma^A_{\bar \ell \nu} = 3 \frac{M_V}{M_A}
\Gamma^V_{\bar \ell \nu}~, 
\qquad\quad
\Gamma^A_{\bar f f} = \frac{2 N_f  \Gamma^A_{\bar \ell\nu} }{c_W^4}  
\left[ ( T_3^f - s_W^2 Q_f )^2 + ( s_W^2 Q_f)^2 \right]~.
\ee

\section{Cross sections}
\label{sect:cross}

\begin{figure}[t]
\begin{center}
\includegraphics[width=6.5cm]{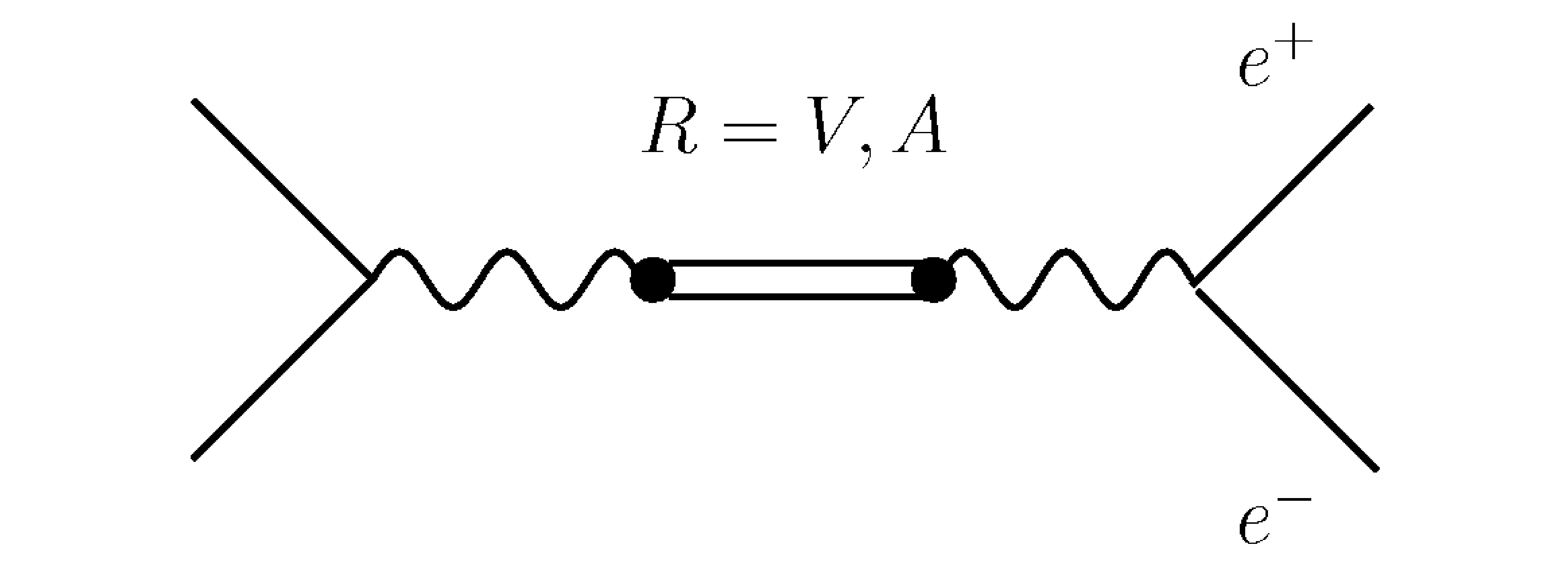}
\hskip 1 cm
\raisebox{0.3 cm}{\includegraphics[width=5.0cm]{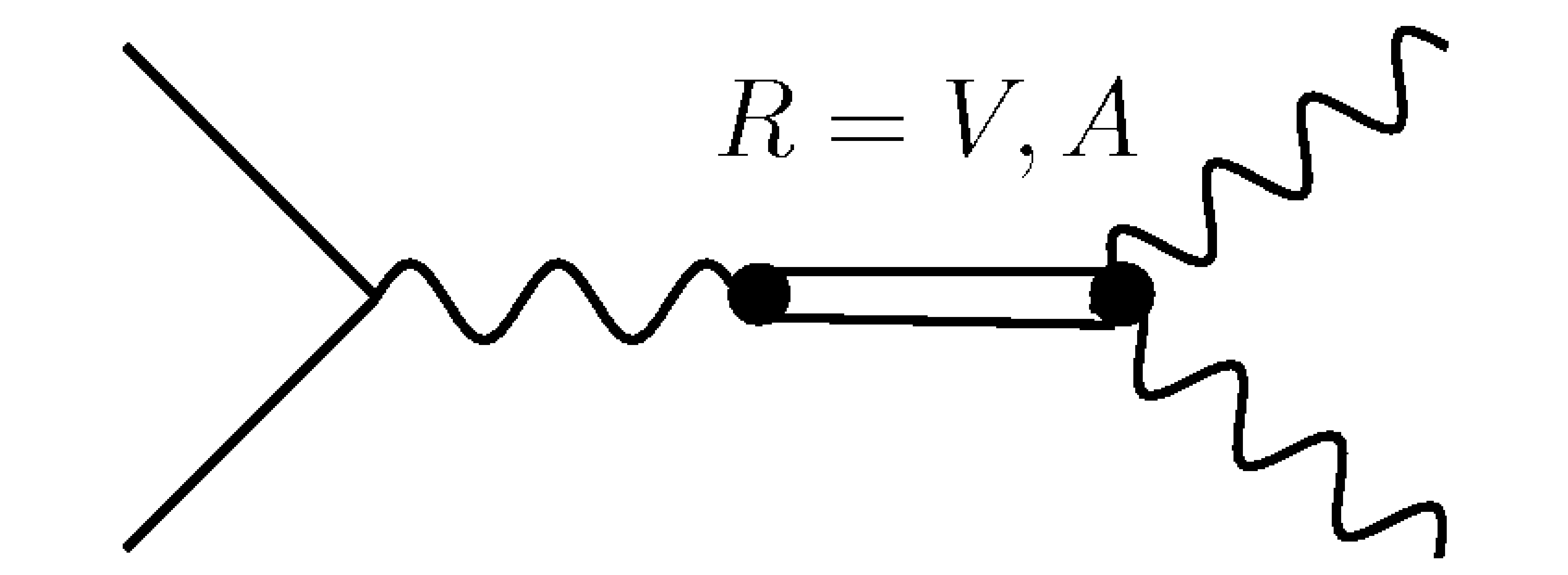}}
\end{center}
\caption{Tree-level diagrams for the resonant production of 
$\ell^+\ell^-$, and $WZ$ (or $WW$) final states (the wiggle lines 
denote generically $W$ and $Z$ states, with both longitudinal
and transverse polarization). \label{fig:1} }
\end{figure}

\begin{figure}[t]
\begin{center}
\includegraphics[width=12.cm]{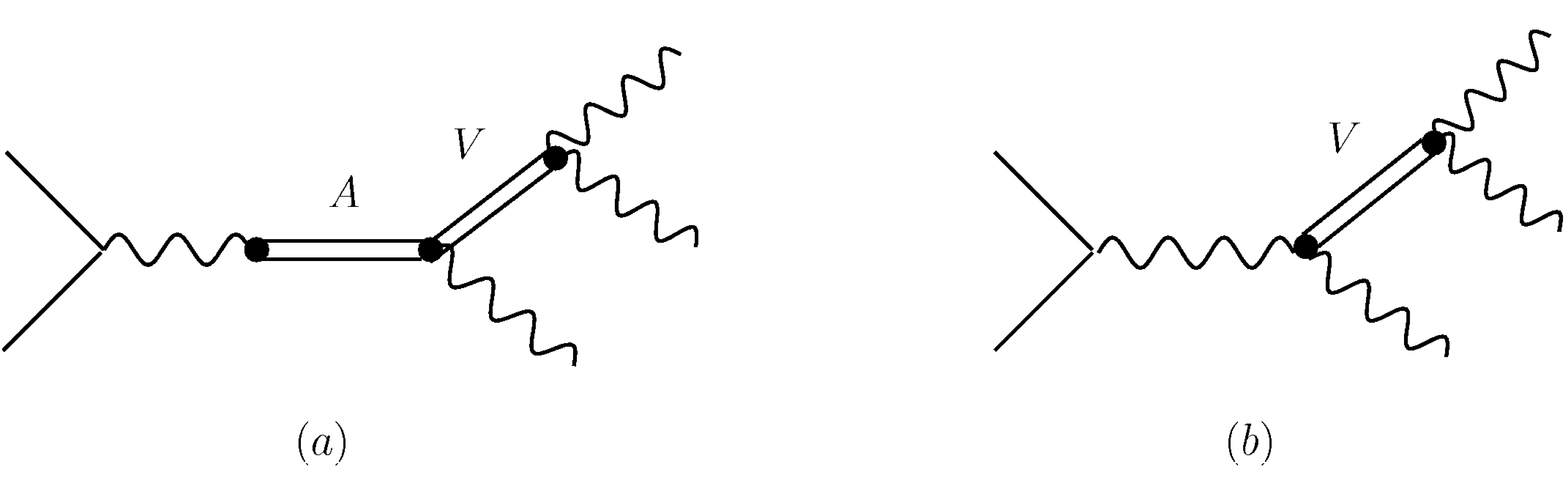}
\vskip -1.0 cm
\end{center}
\caption{Tree-level diagrams for the production of three SM 
gauge fields with a resonant peak in the two-body 
invariant-mass distribution (notations as in Fig.~\ref{fig:1}).
\label{fig:3} }
\end{figure}

In this section we will consider the Drell-Yan production of resonances leading to $\ell^+\ell^-$, $WZ$, and three SM gauge boson ($WWW$, $WWZ$ and $WZZ$) final states. The processes are collected respectively in Fig.~\ref{fig:1} and Fig.~\ref{fig:3}.
At the partonic level,
the Drell-Yan production of the resonances can be written as 
\be
\sigma(q_i \bar q_j \to R \to {f} ) = 
\frac{ 12 \pi \Gamma_R^2 \cB^R_{\rm in} \cB^R_{f} }{ (q^2-M^2_R)^2 +M_R^2 \Gamma_R^2}
\left[ 1 +\cO\left(\frac{q^2-M^2_R}{M^2_R}\right) \right],
\label{eq:BW}
\ee
where $q^2$ is the invariant mass of the generic final state $f$ and
$\cB^R_{\rm in(f)}$ denote the (on-shell) branching ratios
of the resonance into initial and final states.
The expression (\ref{eq:BW}) is valid only for $q^2$ 
close to the resonance peak and neglecting interference 
effects with SM amplitudes. As we will show, the latter is 
a good approximation for the leading two and three SM gauge
boson final states and relatively light spin-1 masses 
(up to $\sim 1$ TeV). On the contrary, interference effects 
cannot be neglected in the purely leptonic final states. 

The convolution with the initial-state parton distribution 
functions is the same occurring in Drell-Yan processes 
within the SM (up to tiny higher-order corrections). 
As a result, it is convenient to normalise the resonant 
cross sections to SM Drell-Yan processes. 
To this purpose we define
\bea
\cF^{R^+}_{f}(q^2) = 
\frac{\sigma(u \bar d \to R^+ \to {f})}
{\sigma(u \bar d \to \mu^+ \nu )_{\rm SM}}~, \qquad 
\cF^{R^0_q}_{f}(q^2) = 
\frac{\sigma(q \bar q \to R^0 \to {f})}
{\sigma(q \bar q \to \mu^+\mu^- )_{\rm SM}}~.
\eea

In the charged case, where we have a single initial state,
the situation is particularly simple: using the form 
factor introduced above the complete $pp$ cross section 
for $q^2 \gg m_{W}^2$ reads
\be
\frac{d}{dq^2} \sigma(pp \to R^+ \to {f} ) 
= \cF^{R^+}_{f}(q^2) 
\frac{d}{dq^2} \sigma(pp \to \mu^+ \nu )_{\rm SM}~,
\ee
where the explicit expression for the form factor is:
\be
\cF^{R^+}_{f}(q^2)  = 
 \frac{12\pi F^2_R \cB^R_{f} }{M_R \Gamma_R} 
\frac{ q^2 \Gamma_R^2 }{ (q^2-M_R)^2 +M_R^2 \Gamma_R^2 }~.
\ee 
For instance, in the $V\to WZ$ case 
this implies
\be
\cF^{V^+}_{WZ}(q^2) 
\approx 80 \times \left(\frac{1~{\rm TeV}}{M_V}\right)^4 \left(\frac{F_V}{2 G_V} \right)^2
\frac{q^2 \Gamma_V^2 }{ (q^2-M_V)^2 +M_V^2 \Gamma_V^2 }~,
\label{eq:exWZ}
\ee
with 
\be
\Gamma_{V^+} \approx (36~{\rm GeV})\times \left(\frac{ 3 G_{V}^{2}}{v^2}\right) 
\left(\frac{M_V}{1~{\rm TeV}}\right)^3.
\ee
The large numerical coefficient in (\ref{eq:exWZ}) and the narrow vector decay 
width allow us to safely neglect the interference with 
the SM in this channel, at least up to $M_V \sim 1$~TeV.

In the neutral case the situation is slightly more 
complicated by the presence of two basic independent 
partonic processes. The complete $pp$ cross section 
for $q^2 \gg m_{Z}^2$ can be written as
\be
\frac{d}{dq^2} \sigma(pp \to R^0 \to {f} ) 
= \cF^{R^0_u}_{f}(q^2) 
\frac{d}{dq^2} \sigma_{(u)}(pp \to \mu^+ \mu^- )_{\rm SM}~+
\cF^{R^0_d}_{f}(q^2) 
\frac{d}{dq^2} \sigma_{(d)}(pp \to \mu^+ \mu^- )_{\rm SM}~,
\ee
where $\sigma_{(q)}(pp \to \mu^+ \nu )$ denote the
SM Drell-Yan cross section induced by the partonic 
process $q\bar q  \to \mu^+ \nu$. The explicit expressions for the neutral 
form factors are:
\be
\cF^{R^0_q}_{f}(q^2)  = 
 \frac{24\pi F^2_R N_q^R \cB^R_{f} }{M^3_R \Gamma_R} 
\frac{ q^4 \Gamma_R^2 }{ (q^2-M_R)^2 +M_R^2 \Gamma_R^2 }~,
\ee
where 
\bea
&& N_q^V = C_q^{-1} \left\{ \left[ (1-2s_W^2) T_3^q +s_W^2 Q_q \right]^2 
+ (s_W^2 Q_q)^2 \right\}, \quad 
N_q^A = C_q^{-1}\left[ ( T_3^q - s_W^2 Q_q )^2 + ( s_W^2 Q_q)^2 \right],  \no \\
&& C_q = \left[ ( T_3^q -  s^2_W Q_q )^2 + ( s_W^2 Q_q)^2 \right]
\left[1 - 4 s_W^2 + 8s_W^4\right] +16 s_W^4c_W^4 Q_q^2 \no \\
&& \qquad\quad +4s_W^2c_W^2Q_q(T_3^q-2s_W^2Q_q)(1-2s_W^2)~. 
\eea
Numerically, $N_u^V \approx 0.67$, 
$N_d^V \approx 0.81$, $N_u^A \approx 0.47$,
$N_d^A \approx 1.19$.

\subsection{The three gauge boson final state}
\label{sect:3Wth}

The final state with three SM gauge bosons ($WWW$, $WWZ$, $WZZ$) 
is dominated by the  $V W(Z)$ intermediate state, with subsequent
decay of the heavy vector into a pair of longitudinal gauge
bosons. The relevant diagrams are shown in Fig.~\ref{fig:3}. 
In this case the pure electroweak SM amplitude is 
totally negligible, while the relative weight of the 
resonant $A \to V W(Z)$ processes compared to the non-resonant
$V W(Z)$ production depends on the (unknown) mass spectrum of the 
heavy spin-1 fields. The general structure of the amplitude, 
taking into account also the non-resonant $V W(Z)$ production, 
can be written as 
\be
\sigma(q_i \bar q_j  \to V W ) = 
\frac{ 12 \pi \Gamma_A^2 \cB^A_{\rm qq} \cB^A_{VW} }{ (q^2-M^2_A)^2 +M_A^2 \Gamma_A^2}
\left[ K_{\rm res}(q^2) + \frac{(q^2-M_A^{2})^2}{F_A^2} \Delta_{V\pi}(q^2) \right],
\label{eq:3W1}
\ee
where $K_{\rm res}(q^2)$ takes into account the 
off-shell $A$ exchange [$K_{\rm res} (M_A^2)= 1$]
and  $\Delta_{V\pi}(q^2)$ arises by 
the non-resonant $V W$ production. Their explicit expressions are 
\bea
K_{\rm res}(q^2) &=& \left(\frac{M_A^6}{(q^2)^3}\right)\frac{(q^2-M_V^2)^3}{(M_A^2-M_V^2)^3}\left[\frac{g_V^2(M_V^2+2q^2)q^2+6g_Vg_AM_V^2 q^2+g_A^2(q^2+2M_V^2)M_V^2}{g_V^2(M_V^2+2M_A^2)M_A^2+6g_Vg_AM_V^2 M_A^2+g_A^2(M_A^2+2M_V^2)M_V^2}\right]~,  \nonumber\\
&& \no \\
 \Delta_{V\pi}(q^2) &=& \left(\frac{M_A^6}{(q^2)^3}\right)\frac{(q^2-M_V^2)^3}{(M_A^2-M_V^2)^3}\frac{1}{g_V^2(M_V^2+2M_A^2)M_A^2+6g_Vg_AM_V^2 M_A^2+g_A^2(M_A^2+2M_V^2)M_V^2}\nonumber\\
 && \!\!\!\!\!\!\! 
\left\{G_V^2\frac{M_V^2+2q^2}{q^2}-2 G_VF_V\frac{2M_V^2+q^2}{q^2-M_V^2}+\frac{F_V^2}{2}\frac{M_V^4+4q^2M_V^2+(q^2)^2}{(q^2-M_V^2)^2}+g_A \frac{2F_AM_V^2}{M_A^2-q^2}\times 
\right.\nonumber\\
&& \!\!\!\!\!\!\!
 \left. \times\left(3G_V-F_V\frac{M_V^2+2q^2}{q^2-M_V^2}\right)+g_V \frac{2F_A}{M_A^2-q^2}\left(G_V(M_V^2+2q^2)-F_V\frac{q^2(2M_V^2+q^2)}{q^2-M_V^2}\right)\right\}~. 
\eea
Also in this case the partonic result can easily 
be translated into the hadronic cross section by 
means of the form-factor method discussed before.
It should be noted that there are always 
two kinematically independent combinations 
corresponding to the same three SM gauge boson 
final state [e.g.~$(W^+Z)_V W^-$ and $(W^-Z)_V W^+$].
Given the narrow widths of the heavy vectors, 
to a good approximation we can neglect their 
interference and sum them incoherently.

\section{Numerical analysis}
\label{sect:num}

In this section we perform a numerical analysis of the cross sections discussed above for the 
existing hadron colliders, namely Tevatron 
($p\bar p$ collisions) and the LHC ($p p$ collisions).

We start our analysis with a detailed investigation of 
the $e^+e^-$ final state. Here we use the recent 
Tevatron analysis to set bounds on the parameter space 
of our model. We will also illustrate how this effective approach 
allows us to easily implement the constraints 
from electroweak precision observables.
We then proceed to discuss the expectations for the
LHC in the case of two and three SM gauge boson
final states.

\begin{figure}[t]
\begin{center}
\vskip -1 cm
\includegraphics[width=12cm]{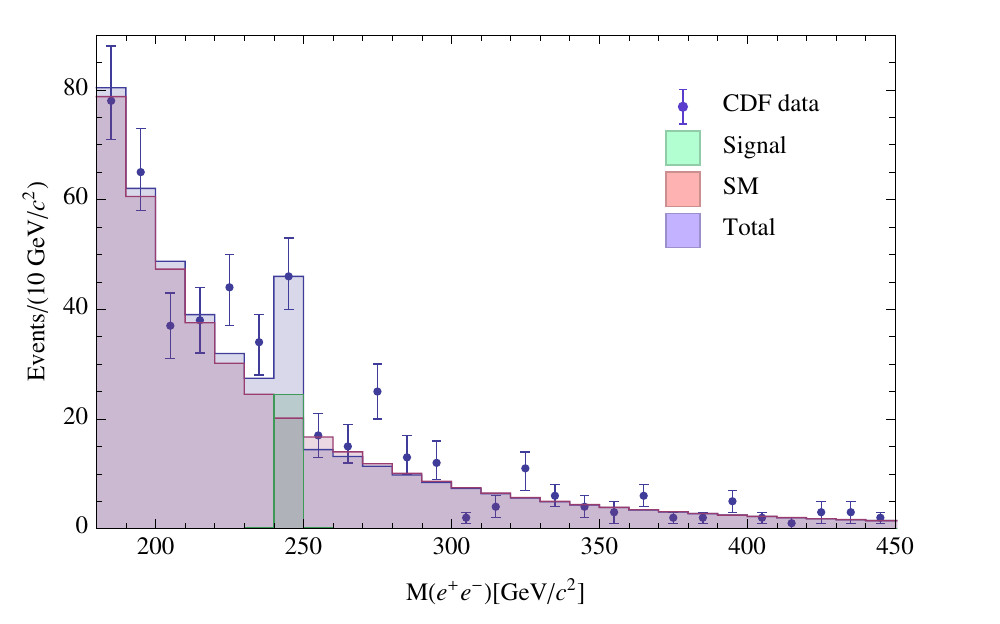}
\end{center}
\vskip -0.5 cm
\caption{High-mass region of the $e^+e^-$ spectrum 
in $p\bar p$ collisions as reported by CDF~\cite{Aaltonen:2008vx}. The {\em signal} and  signal+SM
histograms show the possible contribution of a vector resonance 
of mass $M_V=v$, with $G_V =v/\sqrt{3}$ and $F_V=50$~GeV (see text).}
\label{fig:CDFfit}
\end{figure}

\subsection{The $\ell^+\ell^-$ final state}

The CDF Collaboration has recently reported 
the results of a search for high-mass resonances
in the $e^+e^-$ spectrum of $p\bar p$ collisions at 
1.96~TeV~\cite{Aaltonen:2008vx}. This search does not show 
significant deviations from the SM (with the exception 
of a $2 \sigma$ excess around $250$~GeV, shown 
in Fig.~\ref{fig:CDFfit}) and can be used to set 
limits on the parameter space of our model.

\begin{figure}[t]
\begin{center}
\includegraphics[width=10cm]{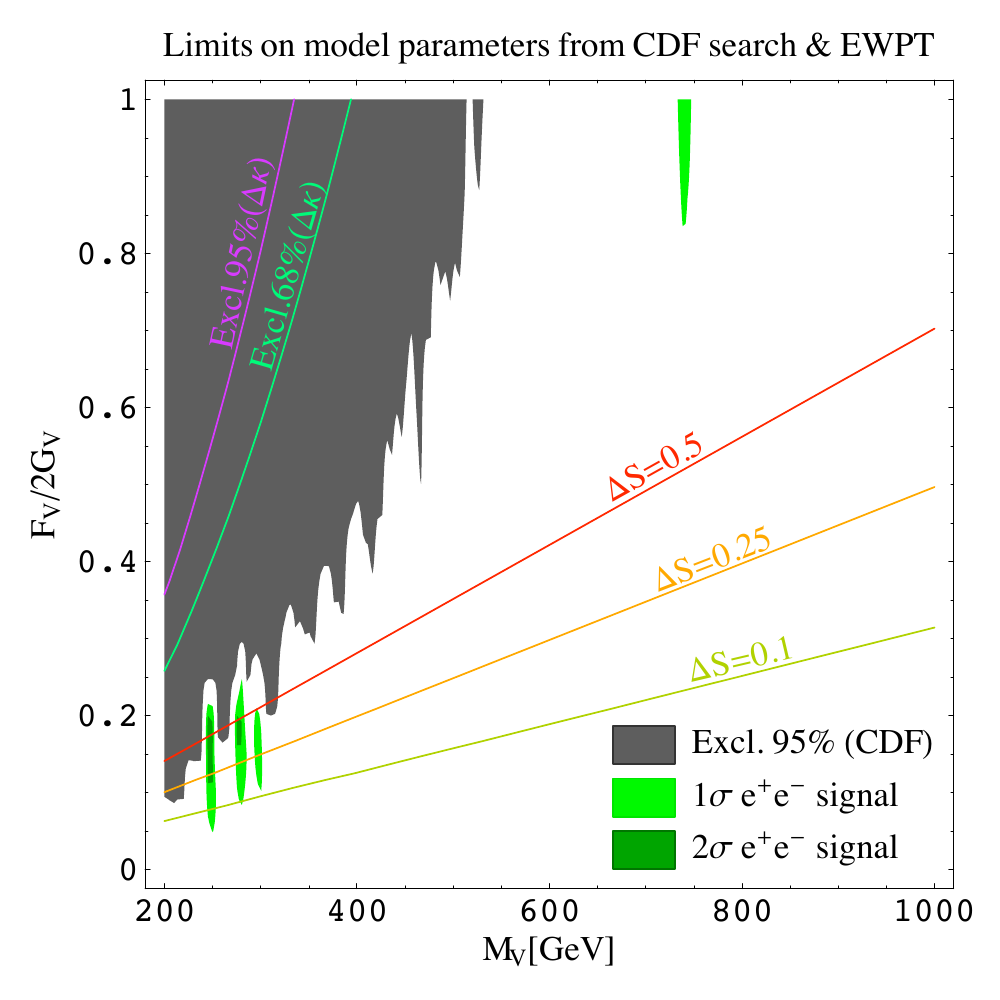}
\end{center}
\vskip -0.5 cm
\caption{Bounds in the $F_V$--$M_V$ plane from $e^+e^-$ CDF data,
and comparison with some of the constraints from EWPO:
$\Delta S$ and anomalous trilinear gauge-boson couplings
(vector contribution only, see text). 
\label{fig:FVMV} }
\end{figure}

Fitting the CDF spectrum with the SM plus a single vector 
resonance leads to the exclusion bounds reported in 
Fig.~\ref{fig:FVMV}. This analysis has been obtained 
fixing $G_V=v/\sqrt{3}$ (from the unitarity constraint), assuming 
$M_A \gg M_V$ (i.e.~neglecting the axial-vector contribution),
and taking into account the interference between the SM and the heavy-vector 
exchange amplitudes. The hadronic cross-section has been
obtained using the form-factor method illustrated 
in Sect.~\ref{sect:cross}, 
normalizing the results to the published CDF data.

The dark gray area in Fig.~\ref{fig:FVMV} denotes the 
region  excluded (at 95\% C.L.) by the CDF data. The small green 
areas correspond to the narrow regions of the parameter space 
where the deviations from the SM are fitted in terms of the 
model parameters. Such deviations are not statistically significant 
yet. However, it is interesting to note that they can be 
fitted in our model, with values of the free parameters which 
are not unrealistic (although they are clearly not the most natural ones).  
In  Fig.~\ref{fig:FVMV} we also report some EWPO bounds, in particular 
the contribution of the heavy vector to the $S$ parameter
\be
\Delta  S =  4 \pi \left( \frac{F_V^2}{M_V^2}- \frac{F_A^2}{M_V^2} \right)\, ,
\ee
and to the trilinear gauge-boson (TGB)  
couplings
\bea
&& \Delta \kappa_\gamma = - \frac{ g^2 F_V G_V}{ 2 M_V^2} 
+ g^2 \frac{\Delta S}{16\pi}~,
\qquad 
\Delta g_1^Z = - \frac{g^2 F_V G_V}{ 4 c_W^2 M_V^2}
- \frac{g^2 s_W^2} {c_W^2(c_W^2-s_W^2)} \frac{\Delta S}{16\pi}~,
\nonumber \\
&& \Delta \kappa_Z = - \frac{g^2(c_W^2 -s_W^2) F_V G_V}{4 c_W^2 M_V^2}
- \frac{2 g^2 s_W^2} {c_W^2-s_W^2} \frac{\Delta S}{16\pi}~.
\label{eq:TGB}
\eea
In the case of $\Delta S$ we show the result for $F_A=0$, 
which can be considered as the maximal 
contribution to $\Delta S$ in this framework. The TGB 
constraints are obtained neglecting the $(\Delta S/16\pi)$
terms in Eq.~(\ref{eq:TGB}), a condition which must be fulfilled in 
any realistic model, given the strong phenomenological constraints
on $\Delta S$.\footnote{~We define trilinear gauge-boson (TGB) 
couplings and oblique parameters ($S$ and $T$) 
as in~\cite{Appelquist:1993ka}. On the TGB, following Ref.~\cite{Dutta:2007st},
we impose the constraints $\Delta \kappa_\gamma =\kappa_\gamma -1 
= +0.16 \pm 0.13$ and $\Delta g_1^Z = g_1^Z-1= -0.09 \pm 0.05$.}

\begin{figure}[t]
\begin{center}
\includegraphics[width=12cm]{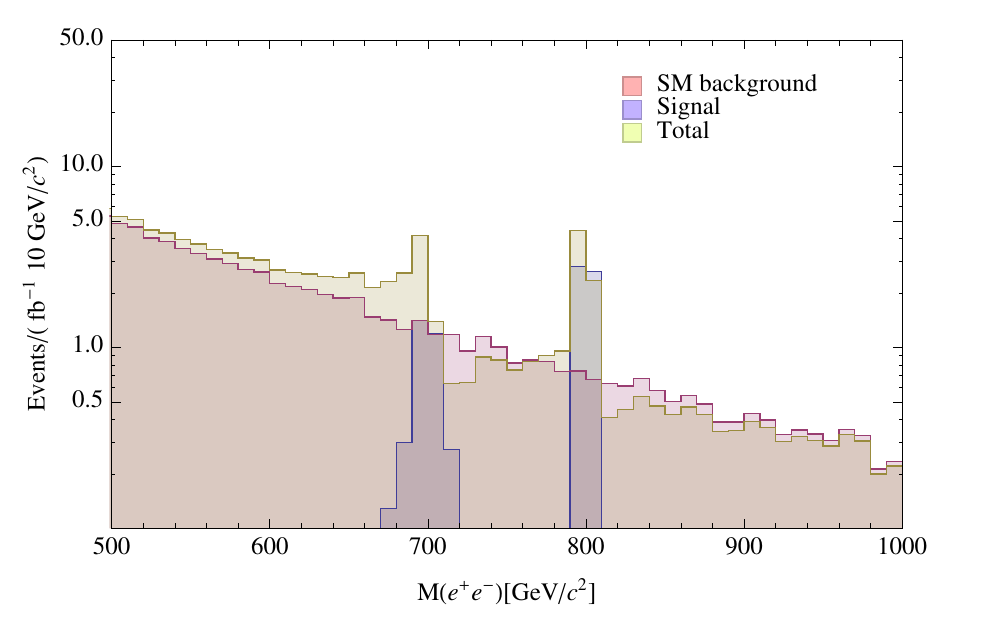}
\end{center}
\caption{Possible signatures of $V$ and $A$ 
states in $pp \to \ell^+\ell^-$ at $\sqrt{s}=14$~TeV. 
The parameters of the model are:
$\{M_V,M_A\}=\{700,800\}$~GeV, $F_V=2G_V=2v/\sqrt{3}$, 
$F_A=300$~GeV (see text for more details). 
The plot does not include any experimental cut and 
reconstruction efficiency.
\label{fig:LHCll} }
\end{figure}

As it can be seen in Fig.~\ref{fig:FVMV}, a light vector resonance could explain the observed 
excess at $m_{e^+e^-} \approx 250$~GeV. This would not be possible
if we impose the hidden-gauge relation $F_V=2 G_V$, but it can 
occur in our more general framework, for $F_V < G_V$. 
Interestingly, the values of $F_V$ and $M_V$ needed to fit 
such excess are not excluded by EWPO. The positive 
contribution to $\Delta S$ need to be partially 
compensated by a negative  $\Delta S$ from the axial 
resonance and, most importantly, by a positive $\Delta T$.
The latter could arise 
at the one-loop level thanks to the mechanism discussed 
in~\cite{Barbieri:2008cc}: imposing that the 
axial resonance acts as a cut-off of the one-loop quadratically 
divergent $\Delta T$ generated by the light vector field, one obtains
\be
\Delta T \approx  \frac{3\pi} {c_{W}^{2}} 
\left( \frac{F_{V}-2G_{V}}{2M_{V}}\right)^{2} \frac{M_A^2 }{16\pi^{2}v^{2}}~.
\ee
Imposing a good EWPO fit we can thus fully determine the properties of both 
$V$ and $A$ states. The axial mass turns out to be 
determined quite precisely, $M_A \approx 1.3$~TeV, 
providing a simple testable prediction of this framework
if the {\em signal} in Fig.~\ref{fig:FVMV} would become 
statistically significant.\footnote{~An even more stringent 
prediction is the appearance of an excess similar 
to the one in  Fig.~\ref{fig:FVMV} in the $p\bar p \to \mu^+\mu^-$ 
spectrum. The absence of significant deviations from the SM 
in the latter~\cite{Aaltonen:2008ah}
reinforce the explanation of Fig.~\ref{fig:FVMV}
in terms of statistical fluctuations only.}

Leaving aside CDF data, in Fig.~\ref{fig:LHCll}
we show typical signatures of $V$ and $A$ states 
in $pp \to \ell^+\ell^-$ at the LHC for more natural 
values of their masses. Here the normalization of 
the cross section and the irreducible SM background 
has been obtained using Madgraph~\cite{Alwall:2007st}. 
In the case of the vector resonance, 
the signal is obtained for $F_V=2G_V$ 
(fixing $G_V$ from unitarity). As it can be seen, even in this 
favourable case (maximal value of $F_V$), the vector peak
is hardly visible if $M_V > 700$~GeV. On the other hand, 
the axial peak could be seen even at larger masses if
$M_A-M_V \ll M_A$, leading to a small axial width (see 
Sect.~\ref{sect:widths}) and a corresponding enhanced peak. 
The axial signal in  Fig.~\ref{fig:LHCll} is obtained 
fixing  $M_A = 800$~GeV, $F_A = 300$~GeV,  
$g_A=1/2$, and  $g_V=0$ 
(the value of $F_A$ follows from the 
requirement of a satisfactory EWPO fit).

\subsection{Two and three gauge boson final states}

\begin{figure}[t]
\begin{center}
\includegraphics[width=12cm]{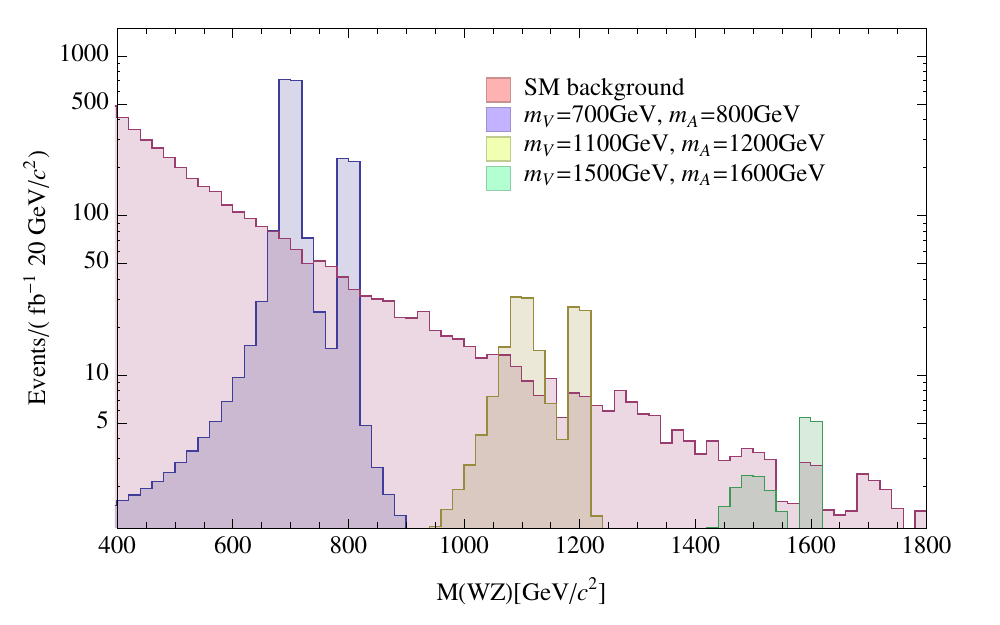}
\end{center}
\caption{Invariant mass spectrum of $WZ$ pairs produced 
in $pp \to WZ$ at $\sqrt{s}=14$~TeV,
with contributions from  $V$ and $A$ states.
All resonance signals have been obtained assuming 
$F_A=F_V=2G_V=2v/\sqrt{3}$, $g_A=1/2$ and $g_V=0$.
The SM background corresponds 
only to the irreducible electroweak production of $WZ$ pairs.
The plot does not include neither experimental cuts nor  
reconstruction efficiencies.
\label{fig:LHCWZ} }
\end{figure}

Since the $R\to ZZ$ channel is forbidden and the $WW$ final state
represents a difficult experimental signature, in the case of 
two gauge-boson final states we focus
our discussion on the $pp \to WZ$ process.
In Fig.~\ref{fig:LHCWZ} we show the typical 
signal at the LHC  for the $WZ$ channel for different 
vector and axial vector masses, assuming a small splitting ($M_A-M_V=100$ GeV). This plot should be used 
only  for illustrative purposes, given that we have not
included the decay branching ratios of 
$W$ and $Z$ bosons and the corresponding 
detection efficiencies. 
As in Fig.~\ref{fig:LHCll}, the vector signal 
is obtained for $F_V=2G_V$ and fixing $G_V$ from unitarity. 
In order to illustrate the possible contribution 
of the axial resonances in this channel, we also 
set $F_A=F_V$, $g_A=1/2$, and  $g_V=0$. 
The normalization of  
the cross section and the irreducible SM background 
has been obtained using Madgraph~\cite{Alwall:2007st}. 

In the three pairs of peaks the first one is always the 
vector signal. As it can be seen, for $F_V=2G_V$ and $M_V=700$~GeV
we should expect $\sim 10^3$~$WZ$ pairs/${\rm fb}^{-1}$,
well above the irreducible SM background, 
or the electroweak cross section for the production 
of $WZ$ pairs in the SM. A realistic evaluation of the 
signal efficiency and the corresponding signal/background 
ratio for this channel is beyond the scope of the present 
analysis. Here we simply note that using the leptonic 
($\ell=e,\mu$) decays of both $W$ and $Z$ 
bosons leads to a theoretical efficiency of 
$\epsilon_Z^{\ell} \times  \epsilon_W^{\ell} \approx 1.5\%$.
For  $F_V=2G_V$ and $M_V=700$~GeV
we should thus expect $\cO(10)$ 
events/${\rm fb}^{-1}$ of rather clean leptonic final states, which 
seems to be a promising signal even with few ${\rm fb}^{-1}$ of
integrated luminosity (similar conclusions 
have indeed been obtained in~\cite{Belyaev:2008yj}, 
where a more accurate simulation of this process
has been presented).

To a large extent the structure of the vector peak in the $WZ$ channel 
is model independent: at fixed $M_V$ it can only be rescaled by the 
ratio $F^2_V/(2G_V)^2$ (the result obtained for $F_V=2G_V$ is what is
expected in the hidden-gauge models). In Table~\ref{Tab:sum} 
we report the corresponding cross-sections both at 
$\sqrt{s}=14$~TeV and $\sqrt{s}=10$~TeV. 
For light masses the expected cross sections are close
to the present sensitivity of Tevatron for $WW$ and $WZ$ 
final states: we have explicitly checked that the
recent measurements/bounds on $WW$~\cite{Abazov:2009ys}
and $WZ$~\cite{Aaltonen:2009hv} final states do not pose 
more stringent bounds on the parameter space of the model 
with respect to those in Fig.~\ref{fig:FVMV}.

\begin{table}[t]
\begin{center}
\begin{tabular}{|c|c|c|c|} \hline 
 & $M=500$~GeV & $M=750$~GeV & $M=1000$~GeV  \\ \hline 
\raisebox{0pt}[15pt][5pt]{$\sigma(pp\to V^+ \to X)_{\sqrt{s}=14~{\rm TeV}}$} &
\raisebox{0pt}[15pt][5pt]{$11$~pb}   & 
\raisebox{0pt}[15pt][5pt]{$1.2$~pb} & 
\raisebox{0pt}[15pt][5pt]{$0.23$~pb} \\ \hline
\raisebox{0pt}[15pt][5pt]{$\sigma(pp\to V^+ \to X)_{\sqrt{s}=10~{\rm TeV}}$} &
\raisebox{0pt}[15pt][5pt]{$6.7$~pb}   & 
\raisebox{0pt}[15pt][5pt]{$0.7$~pb} & 
\raisebox{0pt}[15pt][5pt]{$0.13$~pb} \\ \hline
\end{tabular}
\end{center}
\caption{Summary of the leading-order  cross sections 
for the production of a light charged vector resonance in $pp$ 
collisions at $\sqrt{s}=14$~TeV and $\sqrt{s}=10$~TeV~\label{tab:cross}.
The results are obtained summing over all decay products 
in the mass range $|M_X - M_V | \leq 3 \Gamma_V$, 
setting $F_V=2G_V$, and fixing $G_V$ from unitarity 
($G_V=v/\sqrt{3}$). The results 
for different values of $F_V$ can be obtained scaling 
the figures in the table by $F^2_V/(2G_V)^2$. 
\label{Tab:sum}}
\end{table}

The axial signal in the  $WZ$ channel 
is subject to a larger uncertainty: for $M_A \gg M_V$ it is a 
subleading decay mode and there is no hope to observe it; however, 
for small splitting the axial peak can even exceed the vector one. 
This is clearly illustrated by the three pairs of peaks
in Fig.~\ref{fig:LHCWZ}: increasing both $M_{A}$ and $M_{V}$ 
at fixed $M_A-M_V$ leads to a reduced relative splitting 
which enhances the axial signal. It is worth stressing that 
a small mass splitting between $A$ and $V$ states 
is not an unlikely configuration: it is the most simple 
way to minimize the contribution to $\Delta S$, 
as required by the EWPO.

\begin{figure}[p]
\begin{center}
\includegraphics[width=13.cm]{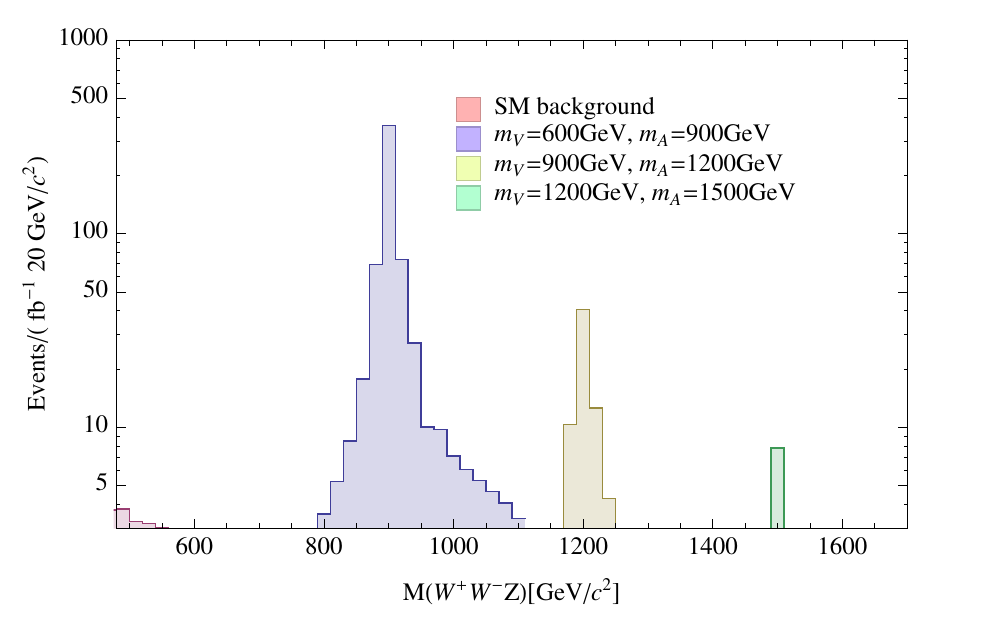}
\end{center}
\caption{$WWZ$ invariant mass distribution in $pp\to WWZ$ at $\sqrt{s}=14$~TeV, 
with contributions from $V$ and $A$ states. Notations and model parameters 
as in Fig.~\ref{fig:LHCWZ}.
\label{fig:LHCWWZ1} }
\end{figure}

\begin{figure}[p]
\begin{center}
\includegraphics[width=8.5cm]{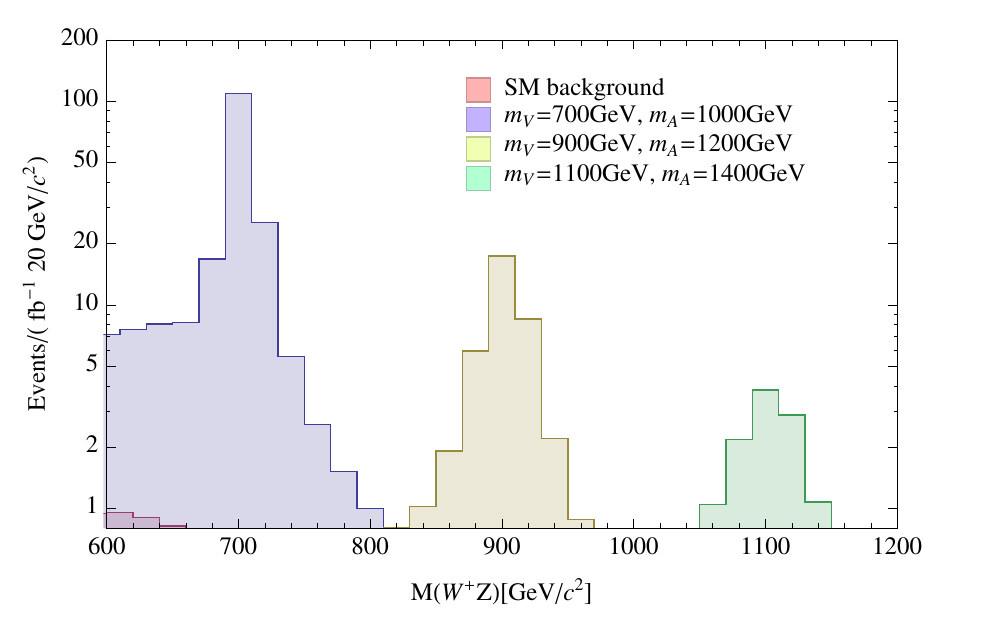}
\includegraphics[width=8.5cm]{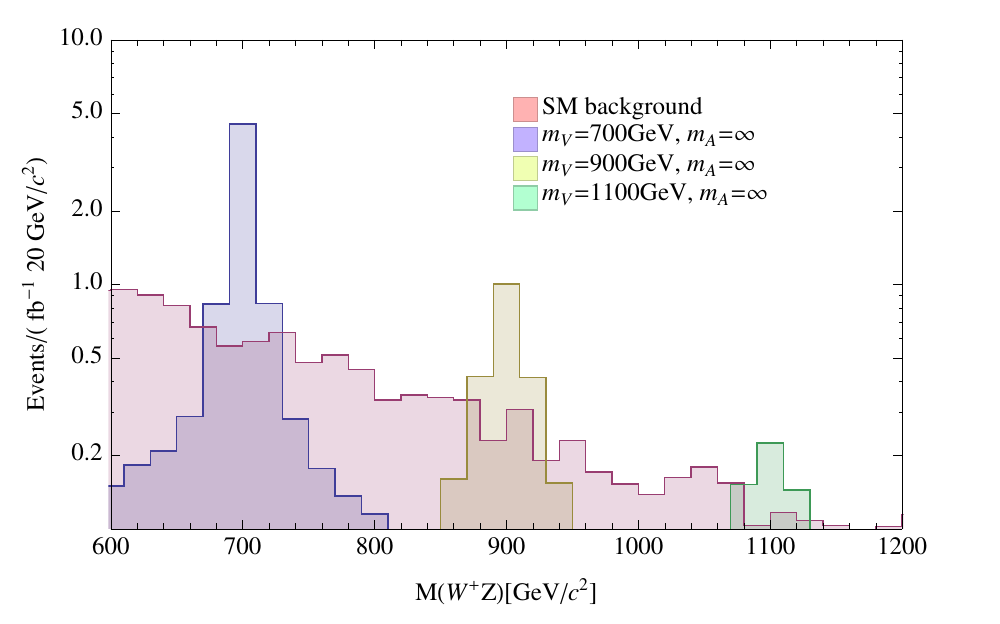}
\end{center}
\caption{$WZ$ invariant mass distribution in $pp\to WWZ$ at $\sqrt{s}=14$~TeV, 
with contributions from both $V$ and $A$ states (left); 
$V$ only (right). Notations and model parameters as in Fig.~\ref{fig:LHCWZ}.
\label{fig:LHCWWZ} }
\end{figure}

\medskip

For large mass splittings the dominant axial decay 
modes are final states with three gauge bosons. 
In Fig.~\ref{fig:LHCWWZ1} we show typical signals 
in the $WWZ$ channel. Almost identical 
distributions are found in the $WWW$ and $WZZ$ 
case. The resonance parameters used for this plots
are the same adopted in Fig.~\ref{fig:LHCWZ} with 
the exception of the masses: a larger mass splitting 
has been chosen in order to increase the signal.
The most interesting observation in this case is 
that the irreducible SM background is totally negligible, 
even for large axial masses. Requiring three leptonic
decays leads to rather small efficiencies, and it does not
necessarily lead to a good rejection of the background 
for processes with more than one neutrino.\footnote{~Requiring at most one neutrino, 
or zero missing mass, has the double advantage of suppressing the non-irreducible
background and allowing the study of the three gauge boson mass spectrum.} 
Generalising 
the results of Ref.~\cite{He:2007ge} on the $WZZ$ case, 
an efficient background rejection should be obtained requiring 
two leptonic decays, of which at least one comes from the $Z$.
According to this strategy, the most promising cases are:
\begin{itemize}
\item{[$WWZ$]} one leptonic $W$, one hadronic $W$, and a leptonic $Z$: 
$\epsilon_W^{\ell}\epsilon_Z^{\ell}\epsilon_W^{\rm had} \approx 0.9\%$;
\item{[$WZZ$]} one leptonic $W$, one leptonic $Z$, and one hadronic $Z$:
$\epsilon_W^{\ell}\epsilon_Z^{\ell}\epsilon_Z^{\rm had} \approx 1\%$.
\item{[$WZZ$]} two leptonic $Z$, and one hadronic $W$:
$(\epsilon_Z^{\ell})^2\epsilon_W^{\rm had} \approx 0.4\%$.
\end{itemize}

Thanks to the non-resonant diagram in Fig.~\ref{fig:3}(b), 
a non-standard signal in the three gauge boson final state 
could also be obtained in the limit $M_A\to \infty$, if
$M_V$ were sufficiently light. In such case we would not observe 
a peak in the three-body mass distribution,
but only in specific two-body projections, as illustrated in 
Fig.~\ref{fig:LHCWWZ}. A detailed discussion of such signal
in the $WZZ$ case has been presented in Ref.~\cite{He:2007ge}
in the context of the so-called three-site Higgsless model
(where there are no light axial-vector resonances).
We agree with the conclusion of Ref.~\cite{He:2007ge}
that in the three-site Higgsless model the $WZZ$
signal is clearly visible at the LHC only with an integrated 
statistics of $\cO(100)~{\rm fb}^{-1}$. On the other hand, 
the comparison of the two plots in Fig.~\ref{fig:LHCWWZ}
shows that in non-minimal models, with relatively light 
axial vectors, the same signal could be enhanced up to 
two orders of magnitude, motivating this search even 
in the early stage of the LHC.

\section{Conclusions}

Spin-1 resonances are a general feature 
of Higgsless models: they are typically 
the lightest non-SM states and play a 
key role in unitarizing the theory 
up to a few TeV. 
The most general signature of these states
(in particular of the vector ones, assuming that 
parity is a good symmetry of the new dynamics), 
is their appearance in $WW$ (or $WZ$) scattering. 
This effect is related to the role played by 
the new states in unitarizing the theory and,
to a large extent, it can be predicted in a 
model-independent way~\cite{Bagger:1993zf}.
The only relevant free parameter is the mass 
of the lightest vector state, which is not fixed 
by the unitarity condition. As shown by recent 
analyses (see e.g.~\cite{He:2007ge}), 
detecting such states in $WW$
scattering at the LHC is not an easy task:
for $M_V\sim 700$~GeV an integrated 
statistics of $\cO(100~{\rm fb}^{-1})$ is needed.

In this paper we have analysed the 
Drell-Yan production (or the production via 
$q\bar q$ fusion) of vector and axial-vector states 
in Higgsless models. 
Contrary to the $WW$ fusion, the Drell-Yan production 
is quite sensitive to the details of the model. We have 
analysed the problem in general terms using the 
effective theory approach proposed in Ref.~\cite{Barbieri:2008cc},
where the relevant properties of the lightest spin-1 states 
are described in terms of a few effective parameters, and the 
constraints of EWPO can easily be implemented.
Despite the faster drop of the signal/background
ratio for rising $M_{V(A)}$, compared to $WW$ fusion, 
we find that in a large fraction of the parameter space 
the Drell-Yan production may yield a rather large 
and clean non-standard signal, even for 
integrated statistics of $\cO(1~{\rm fb}^{-1})$.
In addition to the mass spectrum, the key parameters 
here are the effective couplings $F_{V(A)}$, which 
parametrise the (gauge-invariant) mixing of the new states 
and the SM gauge bosons. Interestingly, the determination 
of these parameters could shed more light on the role 
of the resonances in the EWPO~\cite{Barbieri:2008cc}.

Our main conclusions can be summarised as follows:
\begin{itemize}
\item{} For very light masses ($M_V \lsim 800$~GeV), 
the cleanest signal is the $\ell^+\ell^-$ final state.  
In this channel Tevatron is already providing significant 
constraints in the $F_V$--$M_V$ plane, that we 
have summarised in Fig.~\ref{fig:FVMV}. It is worth stressing that relatively low $M_V$ values are still 
allowed, provided $F_V$ is not maximal: a configuration 
which is not allowed in the simplest Higgsless models, 
but is possible (and even favoured by the EWPO) 
in our more general effective theory approach. \\ 
For large values of $F_V$  ($F_V\approx 2G_V$),
 there are realistic chances to observe
deviations from the SM at the LHC, even with a statistics of 
a few ${\rm fb}^{-1}$. However, in this channel 
the signal/background ratio drops very fast with $M_{V}$:
this implies that it is almost impossible 
to detect a signal for $M_V \gsim 800$~GeV
(even with high statistics).
\item{} 
The $WZ$ final state could offer a wider mass reach, 
for sufficiently high statistics. 
Here the starting point are the total cross-sections 
reported in Table~\ref{Tab:sum}. As shown in Fig.~\ref{fig:LHCWZ}, 
the ratio between signal and irreducible background 
is large even for $M_V \sim 1.2$~TeV. Requiring 
leptonic decays of both $Z$ and $W$, to suppress the 
non-irreducible background, the mass region 
$M_V \gsim 1$~TeV could be explored with an
integrated statistics of $\cO(100~{\rm fb}^{-1})$.
\item{}
The $WZZ$ and $WWZ$ channels are the best channels
to search for the axial-vector resonance, if the latter
is not degenerate in mass with the vector and is not
too heavy. If $F_A$ is large, a configuration which 
is favoured by the EWPO, the resonance signal in 
the $WZZ$ and $WWZ$ channels could be as large as 
in the $WZ$ case (for similar resonance masses), 
and would benefit of a smaller irreducible background. 
\end{itemize}

\section*{Acknowledgments} 
We thank Riccardo Barbieri, Roberto Contino, and Vittorio Del Duca
for interesting discussions. We are also grateful to Barbara Mele 
for drawing our attention to the results of Ref.~\cite{Aaltonen:2008vx}.
This work is  supported by the EU under contract 
MTRN-CT-2006-035482 {\em Flavianet}.


\begin{thebibliography}{99}        
{\footnotesize


\bibitem {Casalbuoni:1985kq}
R.~Casalbuoni, S.~De Curtis, D.~Dominici and R.~Gatto,
Phys.\ Lett.\ B \textbf{155} (1985) 95;   
Nucl.\ Phys.\ B \textbf{282} (1987) 235. 

\bibitem {Csaki:2003dt}
C.~Csaki {\em et al.}, 
Phys.\ Rev.\ D \textbf{69}, 055006 (2004) [arXiv:hep-ph/0305237]; 
Phys.\ Rev.\ Lett.\  {\bf 92} (2004) 101802 [arXiv:hep-ph/0308038].  

\bibitem{Nomura:2003du}
  Y.~Nomura,
  JHEP {\bf 0311} (2003) 050
  [arXiv:hep-ph/0309189].

\bibitem{Barbieri:2003pr}
  R.~Barbieri, A.~Pomarol and R.~Rattazzi,
  Phys.\ Lett.\  B {\bf 591} (2004) 141
  [arXiv:hep-ph/0310285].


\bibitem {Foadi:2003xa}
R.~Foadi, S.~Gopalakrishna and C.~Schmidt,
JHEP \textbf{0403} (2004) 042 [arXiv:hep-ph/0312324]. 


\bibitem {Georgi:2004iy}
H.~Georgi,
Phys.\ Rev.\ D \textbf{71} (2005) 015016 [arXiv:hep-ph/0408067].



\bibitem{Chivukula:2003kq}
  R.~S.~Chivukula, D.~A.~Dicus, H.~J.~He and S.~Nandi,
  Phys.\ Lett.\  B {\bf 562} (2003) 109
  [arXiv:hep-ph/0302263].


\bibitem{Bagger:1993zf}
J.~Bagger~\emph{et al.}, 
  Phys.\ Rev.\  D {\bf 49} (1994) 1246
  [arXiv:hep-ph/9306256].


\bibitem{He:2007ge}
  H.~J.~He {\it et al.},
  Phys.\ Rev.\  D {\bf 78} (2008) 031701
  [arXiv:0708.2588 [hep-ph]].


\bibitem{Accomando:2008jh}
  E.~Accomando, S.~De Curtis, D.~Dominici and L.~Fedeli,
  arXiv:0807.5051 [hep-ph];
  Nuovo Cim.\  {\bf 123B} (2008) 809
  [arXiv:0807.2951 [hep-ph]].


\bibitem{Belyaev:2008yj}
  A.~Belyaev {\em et al.},
  arXiv:0809.0793 [hep-ph].


\bibitem{Barbieri:2008cc}
R.~Barbieri, G.~Isidori, V.~S.~Rychkov and E.~Trincherini,
Phys.\ Rev.\  D {\bf 78} (2008) 036012
[arXiv:0806.1624 [hep-ph]].  

\bibitem{Ecker:1989yg}
G.~Ecker \emph{et al.},
Phys.\ Lett.\ B \textbf{223} (1989) 425;
Nucl.\ Phys.\ B \textbf{321} (1989) 311.


\bibitem{Aaltonen:2008vx}
  T.~Aaltonen {\it et al.}  [CDF Collaboration],
  Phys.\ Rev.\ Lett.\  {\bf 102} (2009) 031801
  [arXiv:0810.2059 [hep-ex]].

\bibitem{Appelquist:1993ka}
  T.~Appelquist and G.~H.~Wu,
  Phys.\ Rev.\  D {\bf 48} (1993) 3235
  [arXiv:hep-ph/9304240].

\bibitem{Dutta:2007st}
  S.~Dutta, K.~Hagiwara, Q.~S.~Yan and K.~Yoshida,
  Nucl.\ Phys.\  B {\bf 790} (2008) 111
  [arXiv:0705.2277 [hep-ph]].

\bibitem{Aaltonen:2008ah}
  T.~Aaltonen {\it et al.}  [CDF Collaboration],
  arXiv:0811.0053 [hep-ex].

\bibitem{Alwall:2007st}
  J.~Alwall {\it et al.},
  JHEP {\bf 0709} (2007) 028
  [arXiv:0706.2334 [hep-ph]],
  http://madgraph.hep.uiuc.edu/


\bibitem{Aaltonen:2009hv}
  T.~Aaltonen {\it et al.}  [The CDF Collaboration],
  arXiv:0903.0814 [hep-ex].

\bibitem{Abazov:2009ys}
  V.~M.~Abazov {\it et al.}  [D0 Collaboration],
  arXiv:0904.0673 [hep-ex].

}
\end{thebibliography}
\end{document}